# Fears and Triggers: A Conceptual Study of Vendor-Supplied Maintenance and Maintenance Deferral of Standard Package Software

**Christopher Savage**
Faculty of Business
University of Wollongong, Australia
Email: cns993@uow.edu.au

**Karlheinz Kautz**
Faculty of Business
University of Wollongong, Australia
Email: kautz@uow.edu.au

**Rodney Clarke**
Faculty of Business
University of Wollongong, Australia
Email: rclarke@uow.edu.au

## Abstract

Enterprises rely more on Information Systems (IS) and software than ever before. However the issue of maintaining a vendor-supplied IS, in particular standard packaged software, has been poorly represented within academic literature. This paper presents a conceptual study that synthesises the current state of research concerning the deferral or execution of vendor-supplied maintenance by the purchasing organisation. Based on a systematic review process that adopts the purchaser's viewpoint, a series of fears and triggers emerge and are captured from the literature. Fears are articulated as reasons for the purchasing organisation deferring the installation of vendor-supplied maintenance, whereas triggers are events that upset the equilibrium of the purchasing organisation's IS or software and require the installation of the vendor-supplied maintenance to proceed. Although prevalent in literature, fears and triggers have not previously been recognised as an area of focus for academic research.

## 1   Introduction

Whether software is viewed as "necessary but troubling" (Ben-Menachem 2008, p.241), "commoditized" (Cusumano 2008, p.23), or providing "increased business benefits" (Gable, Chan & Tan 2001, p.367) there is no question that "both the impact of software on life, and our dependence on software is rapidly increasing" (Xu & Brinkkemper 2007, p.531). Organisations requiring software capability that choose not to develop the capability in-house have the choice of commissioning or outsourcing a unique build, or purchasing the capability (Khoo, Robey & Rao 2011). By purchasing from a vendor, the organisation "benefits [from] generic best practices and advanced functionality supported by vendors' research capabilities" (Maheshwari & Hajnal 2002, p.219). Over time, purchasing this capability has become increasingly attractive (Carney, Hissam & Plakosh 2000) and "once an organization has adopted packaged software, upgrades to newer versions are inevitable" (Khoo, Robey & Rao 2011, p.153)

This paper aims to explore the current state of maintenance deferral literature where it relates to the maintenance of vendor software. For this purpose, "vendor software" is considered to be generic software, pre-created by a third-party organisation for the purpose of sale or licensing. Vendor software includes third-party, commercial-off-the-shelf (COTS) software (Vidger & Dean 1998, Carney, Hissam & Plakosh 2000), product software (Xu & Brinkkemper 2007) or packaged software (Khoo, Robey & Rao 2011). In order to consider the maintenance of this software, maintenance is defined as "the process of modifying a software system or component after delivery to correct faults, improve performance or other attributes, or adapt to a changed environment" (IEEE 1990, p.46). The vendor develops and releases the maintenance, but each purchasing organisation may have to expend significant effort to analyse, test and incorporate the maintenance into the production environment which may lead to the "typical option of 'doing nothing'" (Ng 2001, p.451), "IT's usual preference to 'ride [the current version] out as long as possible'" (Khoo & Robey 2007, p.562) in which "neglect is the inertially easy path" (Horning & Neumann 2008, p.112). This conscious or unconscious decision to postpone or delay the implementation of the vendor-supplied maintenance defines deferral.

No previous literature reviews were identified on the topic of vendor software maintenance deferral. Ben-Menachem's (2008) Y2K-inspired literature review addressing the management of software as assets touches briefly on maintenance but without a focus on maintenance or maintenance deferral. In order to address this gap in the body of knowledge, a systematic review of the literature was conducted, from which a series of fears and triggers emerge and are captured from the literature. Fears are articulated as reasons why purchasing organisations defer the installation of vendor-supplied maintenance, whereas triggers are events that upset the equilibrium of the purchasing organisation's vendor software and require the installation of the vendor-supplied maintenance to proceed. Although prevalent in literature, fears that cause maintenance deferral along with triggers leading to maintenance together have not previously been recognised as an area of focus for academic research.

The remainder of this paper is structured as follows: a background of vendor software maintenance and maintenance deferral within an organisational environment is presented, followed by details of the method utilised to conduct the systematic literature review. Key findings are presented in the results section, followed by concluding remarks that will highlight the knowledge gap that provides an opportunity for further research.

## 2 Background

In discussing the role of maintenance within an engineering business enterprise and positioning the relatively new (1980s) academic discipline of maintenance management, Visser (2002) provides an insight to the origins of maintenance management. From the first origins of maintenance with the creation of tools and structures, items were created robustly and operated to failure. Before World War II, specifically through the industrial revolution, systems became more complex but maintenance apart from routine lubrication remained largely something performed at failure. During World War II the need for operational fighter aircrafts challenged attitudes and created a need for preventative maintenance, maintenance before failure, therefore setting in motion a mainstream function supporting availability. Visser (2002) points to Sherwin (2000) as a source of more detail relating to the history of maintenance. From the information presented within the results of this study, the need for a trigger event in performing maintenance strongly indicates that some IS owners are behaving in a pre-World War II mode of operating-to-failure or, operating-to-obsolesce their software investments. Although a software or IS failure does not exhibit the same visible failure of a physical plant, the clear differentiation between the software system being fit-for-purpose and available for use to being degraded, inoperable or no longer suitable for new requirements is analogous to a failure.

Following the purchase of vendor software, the requirement exists to support the software in order to maximise its operational life because "Systems are nevertheless subject to structural deterioration and obsolescence with age" (Swanson & Dans 2000, p.278). The IEEE (1990) definition of maintenance emphasising modifications after delivery cross-references to Swanson's work within Information Systems (IS) literature and his designations of adaptive, corrective and perfective maintenance (Swanson 1976). By installing vendor software, purchasers will have to "be prepared for managing the impacts of [maintenance]" (Khoo, Robey & Rao 2011, p.167). The actual cost of purchasing the vendor maintenance is not a concern as many vendors employ an annual license agreement whereby maintenance is made available to the purchasing organisation without additional charge (Cusumano 2008).

Importantly to this research, Swanson's inclusive view on maintenance has been adopted where 'maintenance' refers to "all modifications made to an existing application system, including enhancements and extensions" (Swanson & Chapin 1995, p.311). This results in the inclusion of major and minor upgrades, patches and maintenance within the definition of maintenance utilised in this study. The maintenance period is commonly referred to as being the longest phase of the software lifecycle (Carney, Hissam & Plakosh 2000, Vigder & Kark 2006, Abdelmoez, Goseva-Popstojanova & Ammar 2007). However, research is conspicuously absent in academic literature when compared to the "large research literature on the reasons why organisations adopt information technology" (Khoo & Robey 2007, p.556).

Within this study, deferral is treated as a conscious or unconscious decision of the purchasing organisation that postpones or delays a course of action. Implicit within this definition of deferral is that the postponed action will have to be performed at some future time which is consistent with the Oxford English Dictionary (2015) definition of defer "To put off (action, procedure) to some later time; to delay, postpone". Referring to road maintenance, Harvey (2012, p.34) captures the essence of maintenance deferral in any context as "deferring maintenance can be seen as a form of borrowing. Funds are saved in the short-term at the expense of higher outlays in the future".

Deferral may become a critical issue for the purchaser of vendor software when the vendor declares an "end of life" (EOL) date, indicating that further maintenance ceases for this version (Reifer et al. 2003). This forces the business to accept a new risk of using unsupported software, or requires them to perform maintenance that moves to a supported version of software (Khoo & Robey 2007). An example of this phenomenon was the EOL for the Microsoft XP operating system (Microsoft 2014). Subsequent to this EOL of some Microsoft products, organisations are entering multi-million dollar deals with Microsoft to extend support for EOL products. For example, a US Navy agreement for "up to [USD]$30.8 million and extend into 2017" (IDG News Service 2015, p.1) and an Australian federal government departments agreement of "[AUD]$14.4 million to continue support … for another year" (Francis 2015, p.1).

Gartner (2010) reports that the global backlog of overall information technology (IT) and software maintenance activities are estimated to be USD$500 billion in 2010 and may exceed USD$1 trillion by 2015, creating a poorly-understood risk for organisations and a systemic risk for large organisations. This view of unseen risk is supported by Ng, Gable & Chan's (2002) earlier work that identified a growing iceberg of hidden maintenance costs for users of ERP systems. Whereas a civil engineering backlog of maintenance may be visible as wear, rust or decay – this IT and software backlog is a hidden concern. The increasing maintenance backlog on purchasers of vendor software will form a part of this global backlog and justifies more in-depth research in order that theories can be developed and applied to assist practitioners in forecasting and managing this backlog.

## 3  Review Method

The execution of the review and the structure of reporting its results as part of a conceptual study follows the literature review model presented by Webster and Watson (2002) which was selected to ensure that repeatable data gathering and logical analysis support the discussions presented and conclusions drawn. A series of papers by Kitchenham and colleagues (Kitchenham 2004; Kitchenham et al. 2009; Kitchenham & Brereton 2013) provided guidance and finer detail specific to the content required from a systematic review.

In preparation for the systematic review, an unstructured review of publically available literature through a State Library was conducted using the terms "maintenance deferral", "project prioritization" and "project prioritisation". Iterative snowball addition of key words and concepts from the resulting papers created 10 different search terms related to the core concepts defined above. To maximise the scope of literature considered for this review, the formal search was not limited to topic-specific databases, popular publications or peer-reviewed papers. This decision is consistent with the observation that a wide net should be cast in order to consider all published articles in a field (Webster & Watson 2002) and further supported by references in the preparation step where value was added through the discussion of maintenance and deferral relating to vegetation maintenance around power distribution lines (Guggenmoos 2013) and road maintenance planning (Harvey 2012).

The Web of Science™ database (Thomson Reuter 2013, Web of Science 2014) was selected for this review due to the wide cross-discipline nature of the index including the Association for Information Systems' Senior Scholars "basket of 8" journals (Association for Information Systems 2011) and all but two journals from The Financial Times 45 top journal list (The Financial Times 2012). In assessing articles for inclusion in this review, the titles of about 14,900 papers returned through the initial search were evaluated. Papers judged to relate to vendor-supplied maintenance, maintenance deferral in any realm or papers sporting ambiguous or "clever" titles (Dybå & Dingsøyr 2008, p.838) were included for abstract screening to limit false exclusion. Vendor-supplied software maintenance or maintenance deferral in any realm had to be apparent within the screening of abstracts in order for papers to progress to the critical review phase. Through the different screening processes a total of 40 papers were included into the review. The results of this analysis are presented in the following section.

## 4  Results

Analysis of the 40 papers passing the critical review criteria is presented in a concept-centric manner (Webster & Watson 2002). None of the papers set out to address the issue of maintenance deferral by a purchasing organisation.

Maintenance management in the engineering realm, as an academic discipline traces its roots into the 1980s (Visser 2002); software maintenance as a discipline heralds from the work of Swanson (1976). It is therefore surprising that despite the broad search terms, every item that passed critical review, mentioned, referred or alluded to the maintenance deferral problem of vendor software; no papers set out to address this issue directly from the purchaser's perspective.

Through the creation of a concept matrix mapping common themes across multiple papers (Webster & Watson 2002), five key concepts emerged from the literature: (1) an acknowledgement that problems exist when considering vendor-supplied maintenance of vendor software, (2) fear as a driver in behaviour, (3) the occurrence of tipping-points, also identified as triggers, which require vendor-supplied maintenance to be undertaken, (4) the consequences of maintenance deferral, and (5) a lament of scarce research on the topic.

## 4.1 Acknowledgement of the Maintenance Issue

Acknowledgement that adoption of vendor software causes a maintenance problem for the purchasing organisation is a strong emergent theme within the literature critically reviewed (Ng 2001, Ng, Chan & Gable 2001, Ng, Gable & Chan 2002, Khoo & Robey 2007, Horning & Neumann 2008). No papers identified through this literature review expressed a dissenting opinion that vendor software is free from maintenance impacts and considerations. Within this acknowledgement, several specific reasons, labeld fears, and aggravating factors were identified that led to organisational caution when assessing vendor-supplied maintenance before implementing it into production environments.

## 4.2 Fears about Maintenance

In identifying reasons for vendor-supplied maintenance deferral from the literature, a common theme of "fears" was deduced. In almost all cases, the consequences of the fear can be avoided through the deferral of vendor-supplied maintenance, or exercising the "doing nothing" option (Ng 2001, p.451). Although only one paper specifically addressed "fear" towards performing maintenance, these fears were developed from the concept re-occurring throughout literature. Fear is commonly a psychological concept related to an emotion perceived by living entities and not related to social entities such as organisations, it is used here to make a distinction from the risks explicitly mentioned by other authors in relation to deferral.

Ellison and Fudenberg (2000, p. 254) explicitly refer to "fear" in their paper with reference to the introduction of backwards compatibility in Lotus 1-2-3 version 3, stating that "potential purchasers of upgrades were no longer deterred by fears that they would be unable to use their old files".

The fear of losing customisations, configurations, or interfaces was the most recognisable in the literature. It extends beyond the obvious IT/IS-based concerns and into the realm of the user where "Users also create idiosyncratic adaptations and workarounds to overcome limitations in any customised software" (Khoo, Chua & Robey 2011, p.329).

Almost as prevalent was the fear that the application of vendor-supplied maintenance would have a huge cost associated with it (Ng, Gable & Chan 2002). Through the acceptance that purchasing organisations implement vendor-supplied systems such as an Enterprise Resource Planning (ERP) system to gain a commercial advantage (Ng, Chan & Gable 2001), then any planned or unplanned expense in monetary or effort-based terms may detract from this profit-making goal. In some of the limited direct references to deferral across multiple different realms, cuts and limits in maintenance budgets are a common occurrence and the flow-on deferral of maintenance is a direct result (Bausch & Hooven 1977, Hybertson, Ta & Thomas 1997, Reifer et al. 2003). A more general economic downturn may also lead to maintenance being seen as "too costly" (Bloch 2011). Although no papers within this review explicitly attempt to cost a vendor-supplied maintenance project, several examples indicate that the effort is measured in multiples of man-years (Anderson & McAuley 2006, Khoo, Robey & Rao 2011). Costs arise through the plethora of equipment and activities required within the organisation to successfully remove the risk posed by the implementation of the maintenance.

The fear that maintenance to one system will cause or expose a chain reaction of integration updates and backward-compatibility issues as a cascade of maintenance requirements is prevalent in the literature. Examples of this fear included minor inconveniences such as missing device drivers following operating system maintenance during the replacement of printers, faxes and scanners (Khoo, Robey & Rao 2011), through to a case of thirteen linked vendor-supplied systems requiring upgrade (Anderson & McAuley 2006). Within vendor-supplied systems, the fear of cascading maintenance applies also to internal maintenance requirements of the vendor-supplied system. A mandatory maintenance action on one module of a vendor solution may cause issues requiring further maintenance of a separate module, a fear captured by Ng, Gable and Chan (2002).

Recording the fear around training effort and user's learning curves in an upgrade of SAP, Khoo, Chua & Robey (2011, p.332) state "Because SAP upgrades usually involved downtime and training, business users normally preferred to defer an upgrade as long as possible" (Khoo, Chua & Robey 2011, p.332) and quantify the training exposure as "as much as 40 [hours] worth" (Khoo, Chua & Robey 2011,

p.333) per user across potentially thousands of users, and a period of "approximately 3 months" (Khoo & Robey 2007, p.559) before users returned to feeling completely comfortable with the new version of the software. Three months to become comfortable with a new system is supported by Ng, Gable and Chan (2002) who recorded a large increase in user-support requests following introduction of a major change to the system in their case study. Mukherji, Rajagopalan and Tanniru (2006) observe through mathematical simulation that organisations with higher change-management costs specifically including training are more likely to defer maintenance.

The fear that a vendor-supplied maintenance release will be of poor quality and introduce bugs and conflicts between existing and new IS resources appears in early work (Lientz & Swanson 1981) and is reconfirmed by authors such as Khoo, Chua and Robey (2011) and Khoo, Robey and Rao (2011). Bachwani et al. (2014, p. 10) open their paper with a comment that "Unfortunately, many of these [software] upgrades either fail or misbehave", a view supported by Arora et al. (2010a) in their description of user's views on the quality of vendor patches being poor.

The fears that a vendor-supplied maintenance release will consume a tremendous amount of effort to analyse, test, or implement and perform are significant and justified. The need for testing is not eliminated through the implementation of vendor software, "its nature shifts from white box (using knowledge of the source code and design) to black box (without knowledge of the source code or design), and system-level testing receives increased attention" (Brownsword, Oberndorf & Sledge 2000, p.52). Khoo, Chua and Robey (2011) record that the effort for their case study of a SAP upgrade took six months. In Khoo and Robey (2007), a Microsoft Windows upgrade is recorded as taking longer than a year to implement across 125 locations.

The unpredictable behaviour of vendors, where "it is difficult to determine when the software will be released" (Xu & Brinkkemper 2007, p.533), also more simply, the "burdensome … rate of change" for vendor software (Carney, Hissam & Plakosh 2000, p.362) leads to fear concerning the inconvenient rate and time of arrival of maintenance releases. Anderson and McAulery (2006, p. 209) voiced a similar concern where "generic patch and upgrade schedules" led the organisation in their case study to focus on continual evaluation of maintenance and rapid adoption only when needed. A number of studies included the fear that maintenance will also disturb the existing equilibrium of IS in the organisation (Brownsword, Oberndorf & Sledge 2000, Carney, Hissam & Plakosh 2000, Anderson & McAulery 2006, Horning & Neumann 2008, Khoo, Chua & Robey 2011, Khoo, Robey & Rao 2011). A fear that a maintenance project can be as difficult and complex as the original installation is indicated in the studies of Hybertson, Ta and Thomas (1997), Carney, Hissam and Plakosh (2000), Gable, Chan and Tan (2001), Khoo, Chua and Robey (2011), and Khoo, Robey and Rao (2011).

Related to the unpredictable behaviour of vendors, where "it is difficult to determine … which features the software will have … or the quality of the resulting software" (Xu & Brinkkemper 2007, p.533), is the fear of the unforeseen, and the impossibility of fully testing a maintenance release and its un-assessable impacts and side effects. The fear of the unknown is implicitly common to all of the fears listed here, but there is a specific fear of the unforeseen; that even when everything is assessed and mitigated, something might go wrong. Although specifically mentioning almost all of the fears identified, Khoo, Robey & Rao (2011, p.164) give a concrete example of this unforeseen fear when pointing to "Unexpected problems with file sharing in Access" in their case study.

One of the most graphic examples of maintenance disrupting an organisation was the failure of a new feature in upgraded software causing "a mess for about three weeks" and one interviewee recalled people saying "somebody needs to get fired, we're losing millions of dollars a day" (Khoo, Robey & Rao 2011, p.161). Resolution of this issue required vendor support and internal organisational changes. A second example of organisational disruption alluded to in Khoo, Chua and Robey (2011) saw a separate company undergoing slowdown in performance and system lockouts subsequent to a three-day outage to perform the upgrade. Interspersed within company-wide issues, individual users reported "files missing … gone to la-la-land" (Khoo, Robey & Rao 2011, p.162) a side-effect expected from the upgrade being described.

The need to test incoming vendor maintenance was unquestioned within literature. Some papers lamented the complications and costs of maintaining environments for testing (Lientz & Swanson, 1981, Hybertson, Ta & Thomas 1997, Carney, Hissam & Plakosh 2000). Others, such as Ng, Gable and Chan (2002) record that their case study target organisation maintained three separate environments, development, test, and production, in order to manage and maintain their vendor-supplied system. In this organisation "All maintenance is done on the development environment first, and then tested in testing environment … In some cases … iterated through several times before the change reaches the production system" (Ng, Gable & Chan 2002, p.94). This observation of iterative testing adds further support to the fear of the test effort required in vendor-supplied software maintenance.

Because the organisation needs the vendor for the support and maintenance of vendor software, the organisation must rely on the claims of the vendor concerning the suitability of the maintenance release as pointed out by Carney, Hissam and Plakosh (2000), Anderson and McAuley (2006), Vigder and Kark (2006), and Bachwani et al. (2014). There is a fear that this leads to and requires a strong dependence on the vendor and trust in the vendor claims. A similar requirement of dependence and a related fear is specifically valid for the documentation that accompanies a release. Such documentation "might be incorrect on incomplete" (Vigder & Kark 2006, p.13). Inevitably, applying maintenance to an operational system may cause conflict with the vendor, as "During the ... testing phase, [IS] staff identified many problems that they attributed to [the] software, but the vendor countered that the problems were related to client [organisation] configuration decisions" (Khoo, Robey & Rao 2011, p.165). A fear of conflict with the vendor can be deduced from this and has been confirmed by Gable, Chan and Tan (2001), Anderson and McAuley (2006) and Aroraet al. (2010b).

Khoo, Chua and Robey (2011) as well Khoo, Robey and Rao (2011) in their case study investigating upgrading of packaged software point to fears regarding maintenance leading to resistance and a user revolt caused by changed software, but also to fears concerning additional work for expert users in the form of training other employees. Finally, Carney, Hissam and Plakosh (2000) hint at a fear that upgraded software might require a re-certification for a certified system.

In summary Table 1 presents the most common concepts categorised as fears expressed across literature assessed for this conceptual study.

| Top 5 fears that the maintenance will ... | Count of papers |
| --- | --- |
| Lead to loss of existing customisations, configurations or interfaces | 13 |
| Have huge costs | 12 |
| Cause a chain reaction of cascading maintenance, integration updates and backward-compatibility issues | 12 |
| Require training efforts and a user learning curve | 9 |
| Be of poor quality, introduce bugs and conflict between existing and new IS resources | 7 |

*Table 1: Top 5 fears leading to vendor-supplied maintenance deferral by the purchaser*

## 4.3 Trigger Events for Maintenance

An identifiable trigger event immediately preceding the installation of vendor-supplied maintenance is emergent in case-study literature. Mukherji, Rajagopalan and Tanniru (2006, p.1692) concluded that their study supported "the idea that investments in upgrades are best made when the gap between the new technology and the one currently in use reaches a 'critical' threshold". This supports the definition of a trigger event as something that causes this threshold to be reached.

The most often-reported theme within the literature concerning triggering the installation of maintenance was to satisfy the need for increased business benefit. This could be achieved through new functionality and features available within a newer release, which fulfils user requests or their changed requirements for improved performance. The desire for a first-mover advantage may also spur some organisations to install a maintenance offering (Mukherji, Rajagopalan & Tanniru 2006). Other examples of increased business benefits are that maintenance "(1) Improves or enhances the way an organisation does business – to streamline best practice or business process and enhance system integration; (2) Improves or enhances the existing [system] functionality; and/or (3) Could keep an existing version away from vendor-support termination – cost effective" (Ng, Gable & Chan 2002, p.101)

Vendors declaring an end-of-life (EOL) or sunset date for support of a particular version were an oft-referenced trigger for maintenance installation, as "Vendors withdraw support for older versions in order to contain and minimise their own maintenance costs, and to guarantee availability of human resources, skills and services support for clients. Hence, they must focus their maintenance support resources on one or few version(s)" (Ng, Chan & Gable 2001). The adoption of vendor software creates a lock-in situation where "they [purchasing organisations] become dependent on the software vendor to provide them with software functionality and technical support" (Khoo & Robey 2007), which means that a vendor declaring an end to that support represents a significant risk to the purchasing organisation.

Lientz and Swanson (1983) in their early studies on maintenance identified a non-software trigger event, the requirement to move from obsolescent hardware or to upgrade the hardware platform to mitigate hardware availability and support issues. This has been confirmed more recently with Anderson and McAuley (2006) in their case study noticeably illustrating the fear of cascading upgrades as an operating system upgrade was required to support the new hardware, which in turn triggered thirteen separate vendor-supplied software upgrades to re-establish and stabilise the information systems on the new hardware. Iannone et al. (2014) have additionally confirmed this trigger. Hybertson, Ta and Thomas's (1997) work was the first of several studies (see Swanson & Dans 2000; Ng, Chan & Gable 2001; Ng, Gable & Chan 2002; Ben-Menachem 2008; Khoo, Chua & Robey 2011) to identify the resolution of an error relevant to the purchaser as a trigger for maintenance.

Organisational policy may aim to identify the occurrence of a tipping point, however apparently contradictory policies with the same aim were identified in separate studies. Khoo and Robey (2007, p.560) capture a policy within their case study requiring the company remain within vendor-support version requirements to ensure "continuous system operation and timely receipt of vendor support if a problem occurred" therefore reducing operational risk. A separate paper identified "The IT policy … was to upgrade every one and a half years. However, due to business changes … the support group had not conducted any upgrades for more than three years" (Khoo, Chua & Robey 2011).

According to Ellison and Fudenberg (2000) a rationale for standardisation as a trigger for maintenance is the need to remain compatible with external parties that interface with an organisation's IS. Khoo and Robey (2007) describe a case where internally in an organisation a need to standardise IS infrastructure which triggers maintenance arises as a consequence of a large business acquisition and merger.

Lientz (1983) recognised a need to remain current with the marketplace as a straightforward trigger for maintenance. This is supported by the studies of Brownsword, Oberndorf and Sledge (2000), Ng, Chan and Gable (2001) and Mukherji, Rajagopalan and Tanniru (2006). Lientz (1983) also noted response to the external environment such as legislation as a trigger for maintenance and Ng, Chan and Gable (2001) further captured through their case study that a vendor had a specific class of maintenance - a legal-change-patch (LCP) that organisations were obliged to implement. These LCP patches were released over time and are sequential pre-requisites for each other and for version upgrades. Although mandatory to apply, Ng, Chan and Gable (2001) observed a scenario where the mandatory maintenance was issued for a modular part of vendor software otherwise not used by this client. Lientz (1983) also first mentioned other environmental factors such as competitive pressure and general social and cultural factors as triggers.

A major social change was identified by Ben-Menachem (2008) as triggering maintenance: the introduction of the Euro currency within the newly constructed European Union. Similarly, an innovative or discontinuous change such as the Internet, move from mainframes to PC, or the introduction of mobile devices could be a trigger requiring maintenance to participate in the new paradigm (Cusumano 2008). Some papers alluded to the vendor maintenance release as a simple and sufficient trigger for a client organisation's reaction to install it (Brownsword, Oberndorf & Sledge 2000, Ng, Chan & Gable 2001, Maheshwari & Hajnal 2002, Ng, Gable & Chan 2002). In this context Ng, Chan and Gable (2001) point however to an organisational management strategy in their case study that grouping-together or bundling any maintenance would have a beneficial impact on the overall cost and benefit of implementing maintenance. Finally, illustrating the causes for maintenance in complex COTS systems, Carney, Hissam and Plakosh (2000) suggest that new exploits or threats that increase the risk in a safety-critical, life-critical or secure system are possible triggers for maintenance which is confirmed by Arora et al.'s (2010a) study.

Khoo and Robey (2007, p. 562) use the concept of motivating forces "to be any event, or requirement that triggers the interest to adopt a newer version of packaged software". However they conclude that even in the case where one or many trigger events exist as motivating forces, the availability of resources within the purchasing organisation is required in order for maintenance to be applied. If the maintenance is not ranked highly against competing priorities, then continued deferral may be the chosen action. Table 2 lists the top 5 concepts identified triggers that lead to maintenance by the purchasing organisation.

| **Top 5 triggers, in which maintenance is required to …** | **Count of papers** |
|---|---|
| Satisfy a need for increased business benefit or exploit first mover advantage | 10 |
| Respond to external environment such legislation, competitive pressures, social and factors cultural | 7 |
| Avoid an end-of-life date where vendor supports stops | 6 |

| Move from obsolescent hardware, upgrade hardware or support new hardware | 6 |
|---|---|
| Resolve an error relevant to the purchaser | 5 |

*Table 2: Top 5 triggers leading to maintenance by the purchasing organisation*

### 4.4 Consequences of Deferral

Carney, Hissam and Plakosh (200) point to deferral as being a logical, considered course of action when the risk of performing the maintenance is calculated to be unacceptable. Bachwani et al. (2014) illustrated an example of unacceptable risk through a vendor-disclosed incompatibility or difficulty between the maintenance item and a specific type of environment. Likewise, if upgrade will expose an incompatibility issue internally, or with external parties, then a considered deferral decision may be made (Ellison & Fudenberg 2000).

The consequences of maintenance deferral can be to avoid expense in the short term, however the legitimacy and suitability of this approach assume that a trigger event will not occur. Should a trigger event occur and be ignored, possible consequences include economic damage to the company (Arora et al. 2010a), higher expenditure and forced outages at a later time (Bloch 2011), or even demise of the purchasing organisation itself (Carney, Hissam & Plakosh 2000). Alternatively, Khoo and Robey reported that a deferral of the "a" and "b" releases was policy within their case study company to avoid stability issues associated with major releases; rather waiting until the stable "c" release, and further allowing time for other large clients of the vendor to upgrade first (Khoo & Robey 2007).

Gartner (2010, p.1) posit that although IT maintenance can be deferred for one to two years, extended periods of deferral can lead to "the application portfolio risks getting dangerously out of date and a "systemic risk" (Gartner 2010, p.1) for large organizations. Khoo and Robey (2007, p. 556) support this view by stating "organizations do not have to upgrade to every new version of software because vendors typically support multiple versions at the same time".

Khoo, Chua and Robey's (2011) work on motivating users to support an upgrade highlighted one consequence of repeated deferral. In spite of the organisational policy being to implement version-upgrade maintenance every one and a half years, they quote from an interview "we have been live on that version for several years so we didn't need a great deal of help from SAP at that point ... The longer you've been on a release, your reliance on the vendor becomes less so your incentive for an upgrade actually becomes less ... We have the choice too." (Khoo, Chua & Robey 2011, p.332). This shows that one possible alternative available following repeated deferral is to completely separate from the vendor's support model and "go it alone" through either maintaining the system in-house, or paying for bespoke support, possibly receiving a lower priority than up-to-date clients of the vendor (Khoo, Chua & Robey 2011). However, the approach of deferring maintenance comes unstuck when vendor-supplied maintenance "that we require urgently" arrives, but has a dependency on a "backlog" of un-installed changes, which occurs because the vendor "seems to assume that you are up to date" (Ng, Gable & Chan 2002, p.100).

### 4.5 Scarce Interest and Research and Calls for Further Research

The literature includes numerous calls for further investigation into the maintenance of vendor-supplied systems, the maintenance deferral phenomenon, and highlights the increasing issue of maintenance backlog in IT systems and infrastructure. Horning and Neumann (2008, p.112) discuss the risks of neglecting infrastructure in general and declare "Chronic neglect of infrastructure maintenance is not a simple problem, and does not have a simple solution". Investigating maintenance management as a neglected dimension of engineering management Visser (2002, p.484) argues that "The discipline of maintenance management will evolve further as engineers, scientists, technicians and managers integrate results from research and practical maintenance operations to build an internationally accepted body of knowledge."

These statements are valid for IT and software infrastructures. As early as 1983 Lientz (1983, p.277) who studied issues in software maintenance requested "much more research is needed in maintenance". In 1995 Swanson stated, "I wouldn't do the same [1979 Dimensions of Software Maintenance] study [today]. I would take a somewhat different tack. I would try to focus on the maintenance of commercial software packages ... Or, I would address maintenance from the user perspective, which has been largely ignored." (Swanson & Chapin 1995, p.307).

Later, Khoo, Chua & Robey (2011, p.334) articulate the limitation of providing just one study and implicitly call for more research with "Although our research provides an initial investigation into the phenomenon of support upgrades, the empirical support for our findings were limited to a single upgrade case." In the same vein Khoo & Robey (2007, p.556) put forward that "... academic research has largely neglected packaged software, with the exception of ERP systems. One of the most neglected issues related to packaged software is the decision to upgrade from one version to another." Likewise Khoo, Robey & Rao (2011, p.167) lament the neglect of investigations into vendor software maintenance issues with "In conclusion, our study unveils many important facets of a relatively neglected phenomenon: the periodic upgrade of a vendor's packaged software application. Given the steady increase in packaged software solutions, organizations need to be prepared for managing the impacts of upgrades on IS staff and users."

In the context of COTS software Hybertson, Ta & Thomas (1997, p.215) state that "COTS use is increasing, and maintenance issues of COTS-intensive systems need to be articulated and addressed." They are supported by Reifer et al. (2003, p.95) who state "Currently, few COTS software lifecycle models address [Component-Based System] maintenance processes".

## 5 Discussion

This study provides a summary of the reasons for maintenance deferral and performance as well arguments presented in the calls for further research on maintenance and additional evidence to support their line of reasoning. It also responds to and advances two research opportunities put forward by Gable, Chan & Tan (2001) in their research framework for "Large packaged application software maintenance" which consisted of 82 research questions. In exploring the rationale for upgrade deferral, the triggers identified refer to the drivers for an upgrade decision as requested by Gable, Chan & Tan (2001) in their question 'What are the drivers behind the upgrade decision?'. The framework also takes up the question 'To what extent can maintenance be avoided through packaged software and hybrid solutions?' Implicit in this question is the assumption that maintenance can be eschewed, which is addressed by the deferral aspect within this study.

The inclusive approach of this conceptual study requires acknowledgement that although derived from published materials, the identified fears and triggers have been distilled from work where they form sometimes an incidental mention during the description of a complementary topic. Though important enough to warrant mentions, often across multiple separate studies and articles, the list cannot be considered complete without further empirical testing.

This study deals with several other challenges. The first challenge within this study was gaining a suitable view of the concept of software. For as long as IT and IS including software investments have existed, there have been attempts to classify the artefacts of investment in a way common to other investments. Through the adoption of a technical vendor viewpoint, Xu and Brinkkemper (2007) divide software into infrastructure, tools or applications, a view somewhat supported within literature although their subsequent definitions of these three classifications are difficult with some items crossing boundaries. Through a Y2K-motivated literature review, Ben-Menachem (2008) developed an alternative understanding and settled on the description of IT, IS and software as most closely representing an asset. This treatment of software as an asset supports references within the study to deferral behaviours including research within the engineering realm and its physical assets.

The next hurdle was the very definition of the key term of "maintenance". Prevalent within the reviewed papers were conflicting definitions of maintenance, patches and upgrades that acted to further confuse attempts to synthesise a clear picture of the topic. This symptom is supported through the diverse search terms required to capture the documents assessed. Within this study the IEEE definition of maintenance was adopted. Although vendor-supplied software maintenance inevitably adds new upgraded functionality, it still fulfils this definition in that the customer judges the maintenance as required in order for the software asset to remain useful. The study is also based on an understanding that the maintenance phase begins following the purchase of vendor software, not following the commissioning and activation of the same. This is because the first maintenance releases "will occur before the system's initial delivery" (Brownsword, Oberndorf & Sledge 2000, p.53)

The definition of deferral has negative connotations – captured in discussing university building maintenance deferral by Kaiser (1980, p.43) as "Defining deferred maintenance is an exercise that often detracts from the more fundamental task of attacking the problem itself. Because of the implication that deferral has been caused by neglect and not by conscious planning, administrators shy away from approaching the main job". This study has demonstrated that deferral has both legitimate and neglect-based causes.

In highlighting the need for different approaches to internal processes caused by the move from a traditional in-house development team to vendor software, Brownsword, Oberndorf and Sledge (2000) provide a lens to understand the vendor-supplied maintenance deferral question. It is possible that some purchasing organisations fail to make the change to utilising COTS-based processes and therefore remain unaware of the "pervasive ramifications" (Brownsword, Oberndorf & Sledge 2000, p.48) both to people and process that are triggered by implementing a vendor-supplied product and subsequent maintenance.

Traditional budgeting sets an IT department's operating budget on an annual basis. Translation of this budget into a staffing allocation extends the assumption of fixed budget into an assumption of staff costs and staff as fixed input, available to perform work. Contained within this work is the effort required to analyse, test, and implement vendor-supplied maintenance into the production environment. However, this study has shown that vendor behaviour when releasing maintenance does not always conform to support such a predictable cycle of resource and budget availability within the purchasing organisation. The case study by Ng, Chan and Gable's (2001) emphasised that if mandatory maintenance was implemented when it arrived from the vendor, 80% of the annual maintenance effort would be consumed through fortnightly implementations; this would result in batching updates into larger, less frequent implementation activities, significantly reducing the total effort required. This is an example of planned and managed maintenance deferral. An implication of this somewhat random vendor behaviour of maintenance production is to introduce a variable requirement for maintenance work, maintenance assessment, preparation, testing, deployment and stabilisation, into an organisation that is geared for a static level of effort. Incorrectly accounting for this variability through the budgeting cycle could logically introduce a financial constraint on the ability of the organisation to implement vendor-supplied maintenance.

When surrendering control of maintenance to the vendor, an organisation largely relinquishes the ability to group, manage or dictate the content of an individual maintenance package and thus either might defer as long as possible or see no need for maintenance. The latter possibility for why deferral occurs is alluded to by Gable, Chan and Tan (2001, p. 358) who ask "To what extent can maintenance be avoided through packaged software and hybrid solutions?" Speculating about the answer to this question, taken to the extreme, organisations may assume that once purchased, no allocation of time or effort is required for ongoing maintenance, and that the problem was solved in the original purchase. This view is reinforced by purchase decisions that fail to build the operational and maintenance costs into the decision process. An alternative outcome from the trigger event may be to re-assess the vendor software and determine that a replacement is necessary. Tan and Mookerjee (2005) present a case for optimally timed system replacement in response to this outcome. Swanson and Dans (2000) also present an examination of the issues leading to a system being retired; again referencing the maintenance cost/effort of the system to support the decision. In this case, a valid approach is to operate the current system without further maintenance, which is an example of conscious deferral.

Arora et al. (2010a) discuss the economic damages to the purchaser if a security threat is not treated by the application of vendor maintenance and put forward that vendor-released maintenance to mitigate a security issue is implemented through a specific process; separate to that of a 'general' maintenance release. As a consequence the behaviours relating to maintenance deferral in the purchasing organisation are not universal and any future investigation of maintenance deferral needs to take context into account.

This conceptual study presents several possible avenues for future research: firstly, empirical research is needed to confirm the existence of the identified fears and triggers. The relations between distinct fears as well as the joint occurrence of triggers and the interplay fears and triggers warrant more research. Beyond confirming and possibly revising the list of fears and triggers for a deeper understanding empirical research thus should extend the investigations of vendor-supplied maintenance deferral behaviour exhibited by purchasing organisations and of trigger events that ultimately end deferral and lead to maintenance to the context and the processes within which these phenomena take place.

## 6  Limitations

Any conceptual study based on a systematic literature review has limitations. This review was performed by a single reviewer, the first author, without an inbuilt peer-review and dispute resolution process performed for the evaluation of each paper against the sequential filtering criteria. This may lead to incorrect exclusion or inclusion of papers. Limiting the initial search to the Web of Science™ database exposed the review to limitations around the Web of Science™ indexing dates and therefore excluded items published before these dates. In order to mitigate this limitation, citations in articles at the completion of filtering were reviewed in order "to determine prior articles you should consider" (Webster & Watson 2002, p.16). In constructing the search terms, there is a limitation using certain operators within Web of Science™, which are less inclusive but allow more flexible matching than a quoted search term.

# 7 Conclusion

This conceptual study provides a first comprehensive review of the literature on maintenance deferral within the area of vendor software. The deferral of vendor-supplied software maintenance has been observed by practitioners but sparsely considered by scholars. A systematic search and filtering process has shown that research in this acknowledged area of concern is scarce. Calls to action are prevalent within the literature and best articulated by Horning and Neumann (2008) as a need for those understanding this issue to educate decision makers. The IS community can inform this process through further research in this area, resulting in theories, frameworks and/or models that will assist practitioners in navigating the increasing maintenance deferral problem in future decades. This review has demonstrated that execution of a broad systematic literature review relating to a sparsely published area of research informs research through the deduction of themes and concepts from an expansive selection of literature. This study supports practice with an understanding of organisational behaviour with regard to maintenance and maintenance deferral. Through an awareness of common fears and triggers, it can help identify deferral causes, develop responses to these fears and forecast the upcoming need for and implementation of maintenance within an organisation.

The concepts of fears and triggers are presented through deduction from literature, with further research needed to validate these concepts within an empirical case-study framework with a conceptualisation of deferral as a process that recognises that deferment and implementation of maintenance take place in a complex social process. Just as "it appears that understanding packaged software upgrade decisions is not advanced very far by considering the traditional factors affecting decisions to adopt technology" (Khoo & Robey 2007, p.556), the differences between the fears leading to the deferral of maintenance and triggers leading to the implementation of maintenance show that understanding the motivations, labeled triggers, that require an upgrade decision in the present, does not explain all of the motivations for deferral, called fears, in the past. Through the analysis of existing studies and calls to action, this study has demonstrated that deferral of vendor-supplied maintenance has been identified as a problem for over three decades, a view little changed from Swanson's 1989 conclusion reflected upon in an interview that "maintenance is still widely misconceived and mispositioned in firms, and that it deserves more management attention" (Swanson & Chapin 1995, p.310). This study has fulfilled the goal of providing a solid foundation for further attention and research in this long neglected area.